\documentclass[11pt,a4paper]{article}

\usepackage[margin=2.6cm]{geometry}
\usepackage{authblk}
\usepackage{amsmath,amssymb}
\usepackage{hyperref}
\usepackage[backend=biber,style=numeric,sorting=none,url=true,doi=true,isbn=false]{biblatex}
\title{Schr\"odinger’s Equation at 100:\\ The Wave Picture That Helped and Possibly Hurt\thanks{Contribution to the web series \href{https://schroedinger100.univie.ac.at/}{``\textit{Schr\"odinger’s Equation at 100}''} of the Faculty of Physics, University of Vienna.}}
\author[1,2]{{\v C}aslav Brukner}
\affil[1]{Institute for Quantum Optics and Quantum Information - Vienna, Austrian Academy of Sciences, Vienna, Austria}
\affil[2]{Faculty of Physics, University of Vienna, Vienna, Austria}
\date{}
\begin{document}
\maketitle

\begin{abstract}
Schr\"odinger’s equation gave early quantum theory a visual language that looked like physics again: a wave evolving by a linear differential equation. This essay argues that the same success also seeded a recurring impulse to keep quantum theory ``classical-looking'' by treating the wave function as a physical wave. Schr\"odinger quickly realized that, for many-particle systems, the wave function is naturally defined on configuration space rather than ordinary physical space, blocking any straightforward reading of it as a literal classical wave. Read through Mach and Boltzmann, who shaped his intellectual outlook most deeply, his achievement appears double-edged: it provided an extraordinarily powerful picture for calculation and discovery, while also warning against taking that picture too literally. I argue that this tension never fully disappeared. It still reappears in modern physics whenever the wave function, or in quantum field theory the field itself, is treated as ontology rather than as part of a representation tied to measurement and observational context, a point sharpened by Bell-type no-go theorems. The centenary moral is: use pictures boldly, but demote them ontologically.
\end{abstract}


\section{The equation that made quantum theory look like physics again}

When quantum theory first solidified in 1925~\cite{Heisenberg1925}, beginning with Heisenberg’s
``Umdeutung'' paper and rapidly taking systematic form through the works of him, Born, and
Jordan~\cite{BornJordan1925,BornHeisenbergJordan1926}, it did so in a language that was powerful
but, to many physicists, nearly uninhabitable: abstract transition amplitudes, noncommuting
matrices, and calculational rules that no longer seemed to resemble any familiar spacetime picture.

Schr\"odinger’s 1926 ``wave mechanics''~\cite{Schrodinger1926Eigen1,Schrodinger1926Relation}
changed that almost immediately. It offered a linear partial differential equation, together with
an object called a ``wave function,'' whose very form invited a way of thinking reminiscent of how
we think about classical waves and fields. It is hard to overstate how much this pictorial turn
accelerated practical work in atomic and molecular physics, and later in quantum chemistry and
solid-state physics. A century later, this very lineage is explicitly celebrated in Vienna in the
centenary symposium ``\emph{The World in One Line---Schr\"odinger’s Equation Turns 100}''~\cite{ESI2026}.

In the simplest one-particle examples, Schr\"odinger’s wave function looks like it lives in ordinary
3D space. That is the source of its seductive clarity: you can draw it, scatter it, let it
interfere, and tell a story that resembles classical wave phenomena. But almost immediately the
theory forces a correction that is not a minor technicality but a deep ontological revision: for an
$N$-particle system the wave function is naturally a function on a $3N$-dimensional configuration
space, not on 3D physical space. The worry was already put sharply to Schr\"odinger by Lorentz in
May 1926: ``\emph{\dots If, however, there are more degrees of freedom, then I cannot interpret the
waves and vibrations physically \dots}''~\cite{Lorentz1926Letter,Barandes2026}.

Schr\"odinger, for his part, repeatedly stressed that the one-electron problem misleads us here,
and he explicitly flagged as the conceptual difficulty for ``\emph{several electrons}'' the fact
that $\psi$ is ``\emph{a function in configuration space, not in real space}''~\cite{Schrodinger1926Relation,Barandes2026}.
Instead, he suggests that ``\emph{the physical meaning}'' belongs not to $\psi$ itself but to a
quadratic function of it, and he explores how one might recover genuinely spatial quantities by
integrating $|\psi|^2$ over all but one particle’s coordinates to obtain effective three-dimensional
electric charge densities. Schr\"odinger’s response is especially revealing for the centenary
because it shows both his realist impulse and where that realism breaks down.

A striking lens on Schr\"odinger’s mature attitude is his 1940 letter to Eddington, discussed in
the IQOQI Vienna ``Bits of History'' note~\cite{IQOQI2021}. In it, he frames his own outlook as
reconciling Mach's and Boltzmann's positions, two influences often treated as philosophically
opposed. From Mach, Schr\"odinger inherits a suspicion of unconstrained metaphysics: concepts
should remain tethered to what can, at least in principle, be connected to experience. From
Boltzmann, he inherits the methodological necessity of \emph{Anschaulichkeit}---models and pictures
as indispensable engines of discovery---paired with the warning that visualizability is not a
certificate of ontology. Taken together, this ``two-key'' stance invites a deliberately
self-critical reading of wave mechanics: the Schr\"odinger equation is an extraordinarily effective
scaffold for predicting and organizing phenomena, but the step from ``useful wave picture'' to
``real wave in ordinary space'' is exactly the kind of move Mach would distrust, and the kind of
move Boltzmann would license only as a heuristic valuable, and even indispensable, but not
automatically ontological.

This is exactly where Schr\"odinger’s achievement becomes double-edged. Schr\"odinger gave us the
best visualization tool in quantum theory, and at the same time a philosophical warning not to
mistake visualization for ontology.

\section{A fork that never quite closed}

The early episode of misunderstanding the meaning of the Schr\"odinger wave function matters because
it exposed a conceptual fork that, in my view, has never fully closed. A century later, one can
still see traces of that unresolved split in how quantum theory is presented, taught, and
philosophically digested.

In his Nobel Prize lecture and related presentations~\cite{Clauser2021LabConfig,ClauserSlides2024},
John Clauser has argued that the literature still oscillates between two textbook
``\emph{schools of thought}'': a laboratory-space presentation that narrates a probability wave
$\psi(\mathbf r,t)$ in 3D, and a configuration-space presentation that takes many-particle states
seriously from the start. He concludes that the lab-space version
``\emph{should never be used \dots\ even if it helps to make Quantum Mechanics easier to understand!}''.
Carlo Rovelli has voiced a related concern, suggesting that Schr\"odinger’s early wave-function
rhetoric pulled the field in an unhelpful direction: ``\emph{In my opinion this misled the field of
physics a bit.}''~\cite{RovelliIrishTimes2018}.

One school of thought treats configuration space primarily as a representational device. In modern
language, the many-particle wave function $\Psi(\mathbf{x}_1,\dots,\mathbf{x}_N)$ is a
\emph{basis-dependent} component representation of an abstract state $|\Psi\rangle$,
\[
\Psi(\mathbf{x}_1,\dots,\mathbf{x}_N)=\langle \mathbf{x}_1,\dots,\mathbf{x}_N|\Psi\rangle,
\]
and, in an operational reading, it is understood as a representation-dependent tool for computing
probabilities in a given measurement context~\cite{Dirac1930,VonNeumann1932}.

The other school of thought doubles down: if the wave function is ``the thing,'' then one must
either (i) restore classical spatiotemporal storytelling by adding additional variables or
trajectories guided by $\psi$ in the lineage from de Broglie to Bohm~\cite{Bohm1952}, or (ii)
preserve pure wave dynamics by treating the quantum state---even the state assigned to the entire
universe, the ``universal wave function''---as the primary ontology, with classical definiteness
emerging only effectively through Everett-style branching into many universes~\cite{Everett1957}.

Historical accounts such as those of Barandes~\cite{Barandes2026} and Allori \textit{et al.}~\cite{Allori2011}
are especially useful here. They show that the founders' discomfort with configuration-space waves
strongly suppressed wave-function realism early on, and that subsequent revivals, especially Bohm's
explicit defense of an ontological wave function, helped to create the conceptual environment in
which the Everett-style wave ontology could look attractive.

In that sense, Schr\"odinger’s equation did not merely advance quantum theory; it also planted a
particularly resilient form of classicalizing reassurance. That was scientifically fruitful, but it
also made it easier, again and again, to hope that the wave function must be \emph{some} real
physical wave. One might be tempted to conclude, as a sociological counterfactual, that this
delayed a fuller break with classical intuitions, a break that later became central to
quantum-information--theoretic thinking about states, measurements, and the meaning of probability,
and ultimately to the emergence of quantum computation, quantum communication, and modern quantum
technologies. But the idea that this break would have come earlier had physicists distanced
themselves sooner from a realist wave picture is too simplistic. The struggle with classical
intuitions was historically productive and unavoidable.

The no-go theorems of the later twentieth century sharpened that lesson. Figures like John Bell~\cite{Bell1964} and Simon Kochen and Ernst Specker~\cite{KochenSpecker1967} were motivated
precisely by the question of whether quantum theory is compatible with a classical notion of reality. In some cases, this work was even driven by the hope of restoring such a picture.
What emerged instead, through no-go results with different assumptions and scope, was that a classical account in which measurement merely reveals pre-existing properties independent of the full \emph{observational context} cannot be maintained. This demonstrated, in a precise sense, that measurement, or more generally observation, is an integral and fundamental part of quantum theory.

At the same time, work on the no-go theorems clarified the distinction between classical and quantum physics, paving the way for later research aimed at harnessing that distinction for quantum information processing. From
that viewpoint, struggling through the failure of classical notions of reality in quantum theory
was necessary historically, and, I would argue, remains necessary for anyone who wants to work in
modern physics. The point, however, is not to remain trapped in that struggle. I believe we have learned the lesson from Bell-type no-go theorems and can move on. Continuing to cling to the remaining possibilities for
retaining the classical worldview now seems to me like adding epicycles upon epicycles.

\section{A centenary verdict that cuts both ways}

At the centenary, it is tempting to say: \emph{the equation succeeded too well}. It restored to
quantum theory a visualizable object evolving by a differential equation. But that success made the
``classical-like'' impulse a continuously renewable source of misleading intuition: ``it’s a wave,
so it must be a wave somewhere.''

That impulse survives, in my view, in technically sophisticated modern forms, especially in the
dominant textbook and calculational culture of quantum field theory (QFT). In the standard operator
formulation of QFT, one is naturally led back to the picture of a fundamental field living not on
space, as in nonrelativistic quantum mechanics, but on spacetime. In the Schr\"odinger picture of
QFT, by contrast, the state is represented not by $\psi(x)$ but by a \emph{wave functional}
$\Psi[\phi(\mathbf{x})]$ over field configurations $\phi(\mathbf{x})$ on space, that is, by an
object on an infinite-dimensional configuration space~\cite{WeinbergQTF1,HatfieldQFT,PeskinSchroeder}.
Formally, this is the many-particle problem amplified: configuration space is no longer
$3N$-dimensional but function-space dimensional. One can treat this as a useful mathematical
representation of an abstract state, but the same formalism has also encouraged the wave functional
to be understood as a fundamental field living on configuration space (``wave-function realism'' or
``configuration-space realism'')~\cite{AlbertWFR,Ney2023WFR,Chen2019RealismWaveFunction,SEP_QFT_Kuhlmann,Myrvold2015WhatIsWavefunction}.
In either formulation, questions of measurement context, localization, apparatus, and
observer-dependence are often treated merely as practical details of how to ``read out'' the
fundamental field. In this regard, it is telling that systematic frameworks for \emph{local}
measurements in QFT have become explicit only relatively recently~\cite{FewsterVerch2020,FewsterVerch2018},
and that, also very recently, in a distinct line of work, the possibility has been emphasized that
incorporating quantum reference frames relative to which QFT is formulated can cure known
singularities~\cite{GiacominiEtAlQRF,HohnEtAlQRF,Witten2022CrossedProduct}.

Going beyond any specific framework, one may say that the root of attempts to leave out the role of
observation in quantum theory lies in the desire to recover a picture of what is going on ``out
there'' when no measurements are performed. In such a reading, quantum-state dynamics is often
imagined as a sequence of sharply defined frames in some configuration space, as if reality were a
film strip of instantaneous states. But if such a sequence is to have \textit{operational}
meaning, quantum theory immediately sets limits on it. Distinguishability is not a matter of how
vividly we can imagine different frames, but of whether any measurement could, even in principle,
tell them apart. In that strict sense, perfectly distinguishable frames would have to correspond to
orthogonal quantum states.

This is precisely \emph{not} what the Schr\"odinger equation gives you. Under unitary evolution,
$|\psi(t)\rangle$ is generically not orthogonal to $|\psi(t+\mathrm{d}t)\rangle$; in fact, for
small $\mathrm{d}t$ the overlap is close to one. Yet our film-like imagination tends to treat ``the
state at time $t$'' and ``the state at time $t+\mathrm{d}t$'' as if they were two distinct,
mutually exclusive snapshots. The mismatch is instructive: the realist movie-picture appears to be
divorced from operational meaning, a metaphysical add-on. In this sense, the pictorial impulse
behind wave-function realism risks abandoning the operational constraints that make quantum theory
empirical. Schr\"odinger’s equation furnishes the most powerful visualization in quantum theory, yet
an operational reading of its dynamics does not support the film-strip ontology that our
visualization instinct tries to read into it.

\section{The unfinished lesson of quantum theory}

The claim of this essay is that the \emph{conceptual lesson} of the configuration-space
obstruction, already visible in 1926, found its natural completion in the 1960s and in later
no-go results that obstruct any observation-independent notion of reality, yet is still not fully
metabolized in parts of the physics community. In quantum information science, the operational
meaning of state assignments, the contextual character of measurement, and the nonclassical
constraints on information processing are treated as core structure and integral to what quantum
theory is telling us about fundamental physics. In contrast, much of the day-to-day textbook and
calculational culture around high-energy theory still treats those notions as optional philosophy
rather than as guidance for theory-building. If that diagnosis is even partly right, then
Schr\"odinger’s wave picture did not merely help physics move forward; it also helped make it
psychologically easier to keep searching for classical-looking comfort precisely where quantum
theory is least classical.

So---did Schr\"odinger’s equation move physics forward? Obviously and magnificently, yes. It
transformed atomic theory, chemistry, condensed matter physics, and much of the computational
culture of modern microphysics. The visualization it brought with it was decisive in making that
success possible. But at one hundred years, it is worth asking whether the impulse to mistake a
useful visualization for an ontology that dispenses with the notion of observation has also made it
harder, even today and across many areas of physics, to absorb the lessons of Bell-type no-go
results. Perhaps, had modern quantum theory earlier resisted that confusion, it might also have
begun earlier to formalize quantum measurements, reference frames, and observational contexts in
quantum field theory, quantum gravity, and cosmology. Some interpretational detours, too, might
then have looked less compulsory. That is, of course, not a complaint about Schr\"odinger’s
achievement. It is a reminder of the double edge of intelligibility.

\printbibliography

\end{document}